\documentclass[twocolumn]{aastex62}

\usepackage{amsmath}
\usepackage{rotating}

\newcommand{\bvec}[1]{{\ensuremath{\bf{#1}}}}
\newcommand{\thetabv}{{\ensuremath{\theta_{BV}}}}

\newcommand{\E}{\ensuremath{{\bf{E}}}}
\newcommand{\Esw}{\ensuremath{{\bf{E}}_{sw}}}
\newcommand{\Esc}{\ensuremath{{\bf{E}}_{sc}}}
\newcommand{\TEsw}{\ensuremath{{\tilde{\bf{E}}_{sw}}}}
\newcommand{\TEsc}{\ensuremath{{\tilde{\bf{E}}_{sc}}}}

\newcommand{\B}{\ensuremath{{\bf{B}}}}
\newcommand{\kv}{\ensuremath{{\bf{k}}}}
\newcommand{\Vsw}{\ensuremath{{\bf{V}}_{sw}}}
\newcommand{\Veff}{\ensuremath{{\bf{V}}_{eff}}}
\newcommand{\TVeff}{\ensuremath{{\tilde{\bf{V}}}_{eff}}}

\newcommand{\va}{\ensuremath{{{V}_{A}}}}

\newcommand{\vsw}{\ensuremath{{{V}}_{sw}}}
\newcommand{\vph}{\ensuremath{{{v}}_{ph}}}
\newcommand{\hkv}{\ensuremath{\hat{\kv}}}

\newcommand{\BO}{\ensuremath{{\bf{B}}_0}}
\newcommand{\oicw}{\ensuremath{{{\omega^{ICW}}}}}
\newcommand{\ofm}{\ensuremath{{{\omega^{FM}}}}}
\newcommand{\kdi}{\ensuremath{{{k_{di}}}}}
\newcommand{\zhat}{\ensuremath{{\bf{\hat{z}}}}}
\begin{document}

\title{The Electromagnetic Signature of Outward Propagating Ion-Scale Waves}

\begin{abstract}

First results from the Parker Solar Probe (PSP) mission have revealed ubiquitous coherent ion-scale waves in the inner heliosphere, which are signatures of kinetic wave-particle interactions and fluid-scale instabilities. However, initial studies of the circularly polarized ion-scale waves observed by PSP have only thoroughly analyzed magnetic field signatures, precluding a determination of solar-wind frame propagation direction and intrinsic wave-polarization. A comprehensive determination of wave-properties requires measurements of both electric and magnetic fields. Here, we use full capabilities of the PSP/FIELDS instrument suite to measure both the electric and magnetic components of circularly polarized waves. Comparing spacecraft frame magnetic field measurements with the Doppler-shifted cold-plasma dispersion relation for parallel transverse waves constrains allowable plasma frame polarizations and wave-vectors. We demonstrate that the Doppler-shifted cold-plasma dispersion has a maximum spacecraft frequency $f_{sc}^{*}$ for which intrinsically right-handed fast-magnetosonic waves (FMWs) propagating sunwards can appear left-handed in the spacecraft frame. Observations of left-handed waves with  $|f|>f_{sc}^{*}$ are uniquely explained by intrinsically left-handed, ion-cyclotron, waves (ICWs). We demonstrate that electric field measurements for waves with $|f|>f_{sc}^{*}$ are consistent with ICWs propagating away from the sun, verifying the measured electric field. Applying the verified electric field measurements to the full distribution of waves suggests that the vast majority of waves propagate away from the sun (in the plasma frame), indicating that the observed population of coherent ion-scale waves contains both intrinsically left and right hand polarized modes.

\end{abstract}

\author[0000-0002-4625-3332]{Trevor  A. Bowen}
\affil{Space Sciences Laboratory, University of California, Berkeley, CA 94720-7450, USA}
\correspondingauthor{Trevor A.~Bowen}
\email{tbowen@berkeley.edu}

\author[0000-0002-1989-3596]{Stuart D. Bale}
\affil{Space Sciences Laboratory, University of California, Berkeley, CA 94720-7450, USA}
\affil{Physics Department, University of California, Berkeley, CA 94720-7300, USA}
\affil{The Blackett Laboratory, Imperial College London, London, SW7 2AZ, UK}
\affil{School of Physics and Astronomy, Queen Mary University of London, London E1 4NS, UK}

\author[0000-0002-0675-7907]{J. W. Bonnell}
\affil{Space Sciences Laboratory, University of California, Berkeley, CA 94720-7450, USA}

\author{Davin Larson}
\affiliation{Space Sciences Laboratory, University of California, Berkeley, CA 94720-7450, USA}

\author{Alfred Mallet}
\affiliation{Space Sciences Laboratory, University of California, Berkeley, CA 94720-7450, USA}

\author{Michael D. McManus}
\affiliation{Space Sciences Laboratory, University of California, Berkeley, CA 94720-7450, USA}
\affiliation{Physics Department, University of California, Berkeley, CA 94720-7300, USA}

\author{Forrest Mozer}
\affiliation{Space Sciences Laboratory, University of California, Berkeley, CA 94720-7450, USA}
\affiliation{Physics Department, University of California, Berkeley, CA 94720-7300, USA}

\author[0000-0002-1573-7457]{Marc Pulupa}
\affiliation{Space Sciences Laboratory, University of California, Berkeley, CA 94720-7450, USA}

\author{Ivan Vasko}
\affiliation{Space Sciences Laboratory, University of California, Berkeley, CA 94720-7450, USA}

\author[0000-0003-1138-652X]{J. L. Verniero}
\affiliation{Space Sciences Laboratory, University of California, Berkeley, CA 94720-7450, USA}

\collaboration{(The PSP/FIELDS and PSP/SWEAP Teams)}
\section{Introduction}

Understanding observational signatures of energy transfer between electromagnetic waves and particle distribution functions will constrain the kinetic processes contributing to heating and acceleration in coronal and solar wind plasma. Instabilities in the solar wind are thought to drive coherent circularly polarized waves at both ion and electron scales  \citep{Gary1992,Gary1993book,Gary2000,Kasper2002,Hellinger2006,PodestaGary2011b,Verscharen2016,Yoon2017,Klein2018,Verscharen2019,Tong2019a,Tong2019b,Verniero2020}. At ion scales, the cold plasma approximation allows two circularly polarized transverse electromagnetic (EM) modes: left-hand polarized ion cyclotron waves (ICWs) and the right-handed polarized fast-magnetosonic waves (FMWs) \citep{Stix1992,Gary1993book}. However, spacecraft observations of transverse waves are often limited to single point magnetic field measurements, precluding determination of intrinsic plasma-frame polarization. Single point magnetic field measurements only estimate the handedness of fluctuations in the spacecraft frame \citep{Narita2009,HowesQuataert2010}. 

In the solar wind, large (supersonic) flow speeds significantly modify spacecraft frame observations of ion-scale electromagnetic waves through Doppler shift \citep{FredricksCoroniti1976,Klein2014}. While the spacecraft frame polarization may be determined through wavelet or Fourier methods, the wave propagation direction in the plasma frame may be difficult to infer from the Doppler-shifted observations \citep{He2011,PodestaGary2011,Klein2014b,Wicks2016,Woodham2019}. Wave vectors of transverse waves are commonly determined using a minimum variance analysis (MVA) of the magnetic field, which gives the direction of minimum variance as the wave-vector propagation direction {\hkv}  \citep{SonnerupCahill,Means1972,Santolik2003,Jian2009,Verniero2020}. However, MVA analysis, and  eigenvector/eigenvalue determinations, cannot distinguish wave propagation direction parallel or anti-parallel the minimum variance direction (i.e. {\hkv} from -{\hkv}) through magnetic field observations alone. Together, the degenerate determination of sunward/anti-sunward propagation direction and the presence of large Doppler shifts preclude knowledge of the plasma frame polarization from single point magnetic field measurements. Rigorous constraint of dynamical processes, which generate and govern the evolution of these waves, inevitably requires an ability to discern FMW from ICW modes as well as the wave-vector propagation direction.

Including electric field measurements significantly increases the feasibility of wave-mode identification \citep{Santolik2003,Bellan2012,Bellan2016}. \cite{Bale2005} use electric field measurements to demonstrate the Alfv\'{e}nic nature of solar wind turbulence. \cite{Salem2012} additionally use the anisotropy of the spacecraft frame measurements of $E/B$ to suggest that solar wind turbulence is consistent with a cascade of quasi-perpendicular Alfv\'{e}nic turbulence. \cite{Stansby2016} use electric field measurements to determine wave-vectors and frequencies in the solar wind frame to provide an empirical determination of the whistler dispersion relation. In the magnetosphere, observations of $E$ and  $B$ have long been used simultaneously to study coherent wave phenomena \citep{Cattell1991,Chaston1998,Chaston2002}.

First results from PSP have revealed the presence of quasi-parallel propagating ion-scale waves with both left and right hand polarizations, which are preferentially observed with radial alignments of the magnetic field \citep{Bale2019,Bowen2020a}. \cite{Verniero2020} show the simultaneous existence of coherent circularly polarized waves with unstable 3D ion-distributions in the PSP/SPAN-ion data. While numerical solutions to warm-plasma dispersion suggest that both FMW and ICW are potentially driven by instabilities, the specific mode composition of the observed waves has yet to be rigorously constrained \citep{Verniero2020}.

The circular polarized events in the PSP data share qualities with wave phenomena observed in the solar wind at 1 AU by the STEREO spacecraft: quasi-parallel propagation, mixed handedness at ion scales, and preference for radially aligned mean fields  \citep{Jian2009}.  Previous inner-heliosphere measurements of coherent ion-scale waves by {\em{MESSENGER}} and {\em{Helios}} revealed significant scaling in the amplitudes and occurrence rates of ion-scale waves \citep{Jian2010,Boardsen2015}.

\cite{Jian2010} propose that populations of left and right-handed waves correspond to ICW modes propagating in opposite directions; the transit time difference between inward (sunward) and outward (anti-sunward) propagating waves are due to solar wind expansion, resulting in different radial scalings for each polarization signature. \cite{Boardsen2015} note that left-handed ion-scale waves are more common than the right-handed waves, and that the amplitude of left-hand waves scales with $\delta B^2 \sim r^{-3}$, consistent with WKB-like propagation suggested by \cite{Hollweg1974}. Meanwhile, the right-handed waves show significantly shallower radial scaling than the left-handed waves \citep{Boardsen2015}.

Coherent waves at ion-scales are generated by instabilities associated with deviations of particle velocity distribution functions from a Maxwellian distribution \citep{Marsch2006}. The generation of left-handed ion-cyclotron waves are commonly associated with a $T_\perp/T_\parallel > 1$ anisotropy, which drives the Alfv\'{e}n/ion cyclotron instability. On the other-hand, the $T_\parallel/T_\perp > 1$ anisotropy typically generates FMW waves by driving the firehose instability \citep{Gary1993book}.  \cite{Woodham2019} assume that waves with left-hand spacecraft frame polarization propagate uniformly outward, and demonstrate agreement with a  $T_\perp/T_\parallel > 1$ anisotropy; their results suggest that the firehose instability, associated with $T_\perp/T_\parallel < 1$, drives intrinsically right-handed waves in-wards \citep{PodestaGary2011b}. \cite{Telloni2019} show strong correlations with the proton temperature anisotropy and polarization, suggesting that damping of MHD turbulence leads to anisotropic distributions, which subsequently emit parallel propagating ICWs.

Temperature anisotropy alone may not produce coherent wave signatures seen in the solar wind \citep{Gary2016,Verniero2020}. The presence of proton (or $\alpha$-particle) beams introduce extra asymmetry into the velocity distribution, and subsequently more free energy, which may drive waves \citep{Gary1991,Marsch1991,Marsch2006,PodestaGary2011}. Differential flow speeds between the beam and the core (as well as density ratios and beam temperature anisotropies) provide additional constraints on the instability thresholds leading to wave generation \citep{Verscharen2013b,Verniero2020}. 

There is significant observational evidence for the impact of the proton beam on wave growth. \cite{Gary2016} demonstrate that theoretical growth rates from observed proton distribution functions may be dominated by the effect of the beam. \cite{Wicks2016} show that the beam drift correlates well with the amplitude of coherent-wave ``storms" at 1 au. \cite{Zhao2019} suggest, based on statistical observations, that both temperature anisotropy and core-beam drift contribute to the growth of left-handed waves at 1 au. In PSP data, \cite{Verniero2020} demonstrate correlations between the presence of a proton beam and significant ion-scale wave activity \citep{Bowen2020a}.

PSP provides electromagnetic measurements through the FIELDS instrument \citep{Bale2016}. Magnetic fields measurements are made through both a low-frequency flux-gate magnetometer and a high frequency search coil magnetometer \citep{Bowen2020b}. Additionally, the Digital FIELDS Board (DFB) samples the four FIELDS antenna in the plane of the spacecraft heat-shield, producing both single ended and differential voltage waveforms at a survey cadence of up to 292.969 Sa/Sec \citep{Malaspina2016}. \cite{Mozer2020} outline the calibration of the electric field $\Esc$ from differential measurements, where a frequency dependent effective length $\alpha(f)$ with $\Esc= \Delta V_{ij}/\alpha$ is constructed using observations of coherent waves at ion-scales and above. However, knowledge of the wave phase speed, and thus a specific wave mode and polarization, is required to perform a precise correction. 

The linearized Faraday equation provides the fundamental relation between {\E} and {\B} for electromagnetic oscillations. In the solar wind plasma frame,

\begin{equation}\Esw=-\vph \hat{\kv}\times\B
\label{eq:em_wave} \end{equation}
and $\vph =\omega/|\kv|$,
such that measurements of $\B$ and the $\Esw$ components perpendicular to $\kv$ uniquely determine both the wave phase speed and propagation vector. However, in a frame moving relative to the solar wind, i.e. the spacecraft frame, the electric field  $\Esc$ includes contribution both from the solar wind electric field {\Esw} and convection:

\begin{equation}\Esc=\Esw -\Vsw \times \B.\end{equation}

For a transverse quasi-parallel electromagnetic wave, the spacecraft frame electric field is given by,
\begin{align}
\Esc=-\vph \hat{\kv}\times\B -\Vsw \times \B\\
\Esc=-(\vph \hat{\kv}+\Vsw) \times \B\\
\Esc=-\Veff \times \B,
\end{align}
with the effective velocity $\Veff$ corresponding to the total wave speed (convected plus phase speed) measured in the spacecraft frame. 

In addition to {in situ} electromagnetic field measurements, PSP makes measurements of the plasma distribution functions with the Solar Wind Electron Alpha and Proton (SWEAP) instrument suite \citep{Kasper2016}. SWEAP can determine the bulk plasma flow $\Vsw$ as well as thermodynamic moments of the plasma, such as density $n$ and temperature $T$. Integrating measurements from FIELDS and SWEAP provide a set of measurements capable of constraining the observed distributions of wave-modes and propagation directions of coherent polarized waves at ion scales.


\section{Data}

Data is obtained from PSP FIELDS and SWEAP on Nov 04, 2018. A continuous wavelet transform, similar to \cite{Bowen2020a}, is used to locate circularly polarized ion-scale events. Magnetic and electric fields are convolved with a set of wavelets $\psi(\xi)$ normalized to unit energy
\begin{equation}
    W_j(s,\tau)=\sum_{i=0}^{N-1} \psi\left(\frac{t_i-\tau}{s}\right)B_j(t_i)\\
   \end{equation}
   with the un-normalized Morlet wavelet  $\psi(\xi)$ defined as
   
\begin{equation}
\psi (\xi)= \pi^{-1/4}e^{-i\omega_0\xi }e^\frac{-\xi^2}{2} \label{eq:wavelet}
\end{equation}
where $\omega_0=6$, with the relationship between wavelet scale and spacecraft frequency approximated as $f\approx\frac{\omega_0}{2\pi s}f_s.$ The local, scale dependent, mean magnetic field is computed using  a Gaussian envelope of each wavelet \citep{Bowen2020a,McManus2020}. The scale dependent mean field determines a local right-handed field aligned coordinate system for each scale $\hat{W}=(\hat{B}_{\perp1},\hat{B}_{\perp2},\hat{B}_0)$.

The polarization relative to {\BO} is computed as
\begin{equation}
\sigma_{B\perp}=\frac{\langle-2\text{Im}(B_{\perp1}B_{\perp2}^*)\rangle}{\langle B_{\perp1}^2+B_{\perp2}^2\rangle}
\end{equation}

where $\langle...\rangle$ denotes time averaging over a period defined by the $e$-folding of the wavelet (i.e. the cone of influence \citep{TorrenceCompo}).

Measurement of a left-hand circularly polarized wave gives $\sigma_{B\perp}=1$, while a right-hand  circularly polarized wave gives $\sigma_{B\perp}=-1$. For each wavelet scale, we identify a polarized event if $|\sigma_{B\perp}|\geq 0.95$, where the start and end times of the event are defined at the transition $|\sigma_{B\perp}|= 0.8$.

The FIELDS electric field measurements are only available in the spacecraft $x$-$y$ plane--where $z$ is approximately sun pointing. Once wave events are identified in the magnetic field, in the mean-field aligned system ($\perp_1$,$\perp_2$,$\parallel$), the subsequent analysis is performed in the spacecraft coordinate system ($x$,$y$,$z$) in order to facilitate comparison with electric field measurements.

 As electric field measurements are limited to the spacecraft $x\text{-}y$ plane, we recompute the magnetic helicity measured along the spacecraft $x$ and $y$ axes, $\sigma_{Bxy}$, for each interval where $\sigma_{B\perp}$ meets the condition for circular polarization.  Differences in  $\sigma_{B\perp}$ between $\sigma_{Bxy}$ result from variations in the angle between the magnetic field and solar wind flow $\theta_{BV}$; accordingly, not all intervals which meet the $\sigma_{B\perp}$ polarization condition maintain helicity in $\sigma_{Bxy}$. To account for this difference, we only consider intervals with at least half the magnetic field measurements with a helicity of $|\sigma_{Bxy}|>0.85$. The electric field polarization $\sigma_{Exy}$ is additionally computed. 
 
\begin{figure*}
    \centering
    \includegraphics{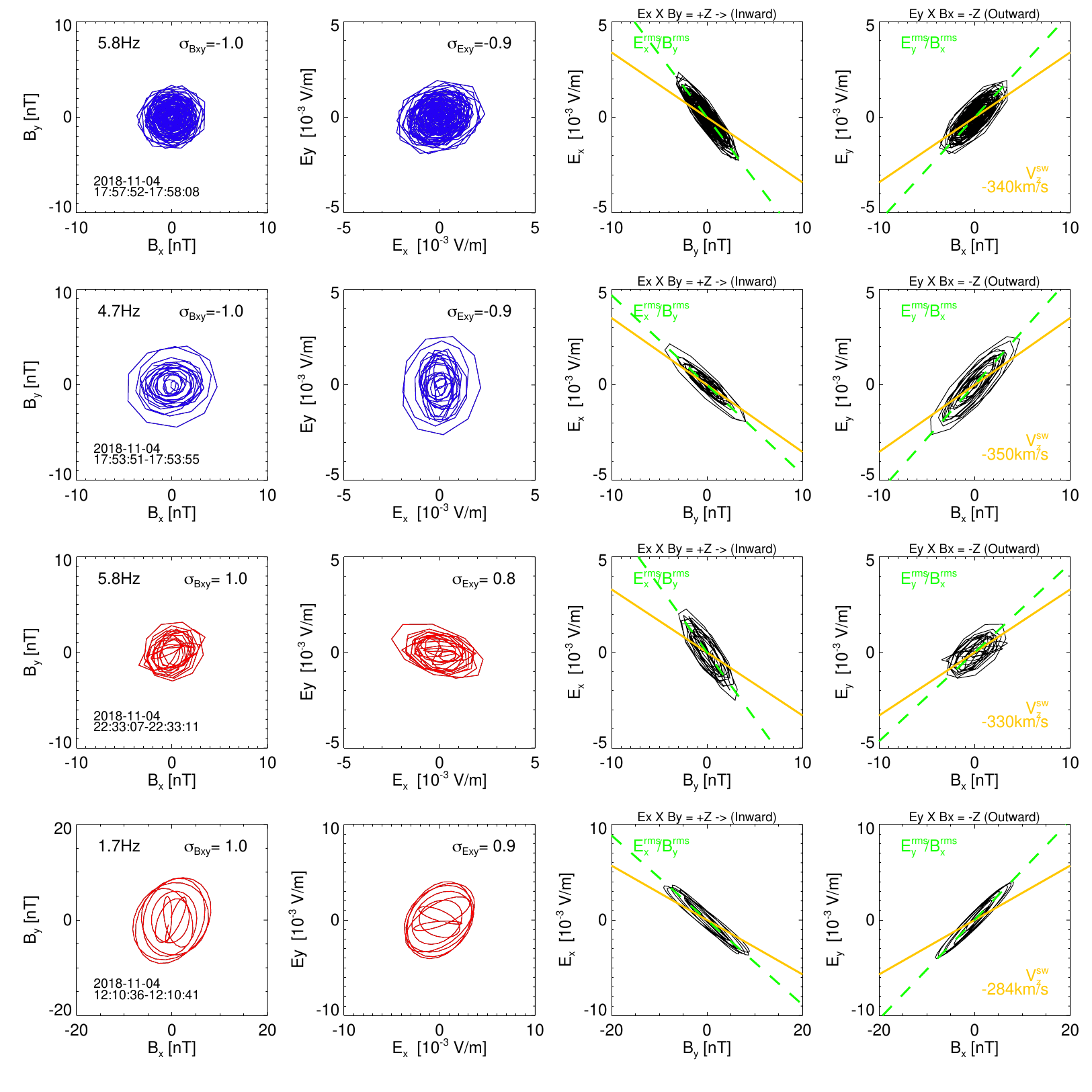}
    \caption{Four example intervals with significant helical signatures. Each row shows one interval. Left-most (first) column shows the hodogram of magnetic fluctuations in $x\text{-}y$ plane. The wavelet frequency and time-span of each event are noted in the first column. Second column shows corresponding hodogram of electric-field fluctuations in $x\text{-}y$ plane. Third column shows phase diagrams of $E_x$ and $B_y$; rms $E_x/B_y$ is shown in a dashed green line, with the total rms $E/B$ shown as solid purple line. Right-most (fourth) column shows phase diagrams of $E_y$ and $B_x$; the measured rms $E_y/B_x$ is shown in a dashed green line, with the total rms $E/B$ in purple. The measured solar wind speed along the $z$-direction is plotted in yellow.}
    \label{fig:f1}
\end{figure*}

 Figure \ref{fig:f1} shows four sample events (one event per row) with helical electromagnetic fields. The first column shows hodograms of the wavelet filtered magnetic field in the $x-y$ plane. The second column shows the corresponding wavelet filtered electric field in the $x-y$ plane. The third and fourth columns show measured $E_x$ vs $B_y$ and $E_y$ vs $B_x$. For transverse circularly polarized waves travelling in the $z$ direction, Equation \ref{eq:em_wave}, indicates that a linear relationship with $E_x/B_y\approx{\vph}_{z}$ and $E_y/B_z\approx-{\vph}_{z}$, where ${\vph}_{z}$ is  in the measured (spacecraft) frame; $v_{ph(x,y)}$ and $B_z$ are taken as small terms. Respective root-mean-square (rms) ratios $E^{rms}_x/B^{rms}_y$, e.g. $E^{rms}_x=\sqrt{\langle{E_x}^2\rangle}$, are plotted, in addition to the line corresponding to $\vph=\vsw$. In these several cases, the measured phase speed of the waves is larger than the local solar wind speed. Additionally the sign of slope of the line indicates outward propagation in the spacecraft frame.

We note that by measuring helicity in the spacecraft coordinate system, we have removed the dependence on $\BO$, which is necessary to determine the plasma-frame polarization. However, during the first perihelion, PSP was connected to a small coronal hole of negative polarity, with a field in the $+\zhat$ spacecraft direction (the solar wind flow is approximately in the $-\zhat$ direction)  \citep{Bale2019,Badman2020}. Since the measurements are taken with a single (positive) background-field direction, no rectification is necessary to account for sector structure associated with heliospheric polarity \citep{Woodham2019,McManus2020,Badman2020}. However, deviations from a $+\zhat$ oriented mean field occur during large-scale magnetic field switchbacks, in which the mean field deflects, and the polarization plane of the quasi-parallel waves is no longer aligned with the $x\text{-}y$ plane: i.e. $\sigma_{Bxy}$ significantly deviates from $\sigma_{B\perp}$ \citep{DudokdeWit2020}. Circularly polarized events occurring within switchbacks are then typically excluded by the condition $\sigma_{Bxy}>0.85$, as the polarization plane is then perpendicular to the spacecraft $z$-axis.


\begin{figure}
    \centering
    \includegraphics{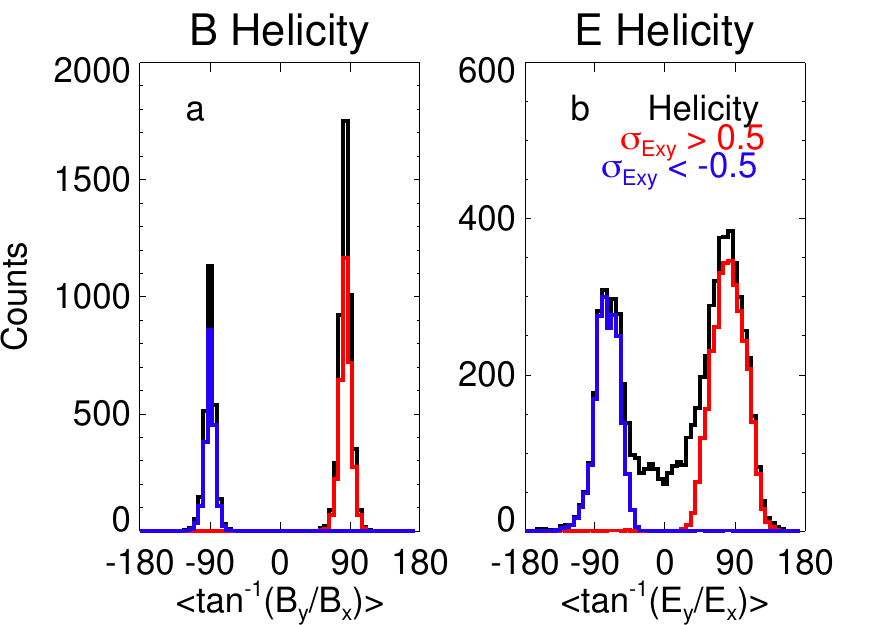}
    \caption{(a) Distribution of phase difference between $B_x$ and $B_y$ for circularly polarized intervals. (b) Distribution of phase difference between $E_x$ and $E_y$ for circularly polarized intervals. Black curve shows the total measured distribution for $\sigma_{Bxy}>0.85$, while the red and blue distributions show the respective conditions of $|\pm\sigma_{Exy}|>0.5$ where $\pm$ corresponds to left/right (red/blue) helicity measured in the spacecraft frame.}
    \label{fig:f2}
\end{figure}

Figure \ref{fig:f2}(a) shows the distribution of measured phase difference between spacecraft $x$ and $y$ magnetic field components for events with  $\sigma_{Bxy}>0.85$. A value of $+90^\circ$ corresponds to left-handed polarization and $-90^\circ$ corresponds to right-handed polarization. Figure \ref{fig:f2}(b) shows the corresponding distribution for electric field measurements. While the electric field measurements are similarly peaked at $\pm 90^\circ$, corresponding to circular polarization, the respective peaks are much broader. There are a significant number of measurements that have helical magnetic fields, but no helical signature in the electric field. This is possibly a result of increased noise in the electric field measurements arising from either the innate non-orthogonality of the sensors, or the optimization routines which generate electric field from the measured differential potentials \citep{Mozer2020}. Additionally, we do not exclude the possibility that physical dynamics other than parallel propagating transverse waves, e.g. an electrostatic component or non-parallel propagation, may be relevant in some cases. We limit our study to measurements with $|\sigma_{Exy}|>0.5$, respectively shown in red (blue) for left (right) handed polarizations, such that our discussion of circular polarization maintains connection with physically realized measurements. In the magnetic field 7153 events with $\sigma_{Bxy}>0.85$ (over all scales) were measured, $|\sigma_{Exy}|>0.5$ is measured in 5191 of these events.

Identifying this subset of events with circular polarization in both magnetic and electric fields provides an empirically observed distribution, which may be compared with theoretical behavior of parallel propagating electromagnetic cold-plasma waves under the effect of Doppler shift.

\section{Dispersion \& Doppler Shift}
\begin{figure}
    \centering
    \includegraphics{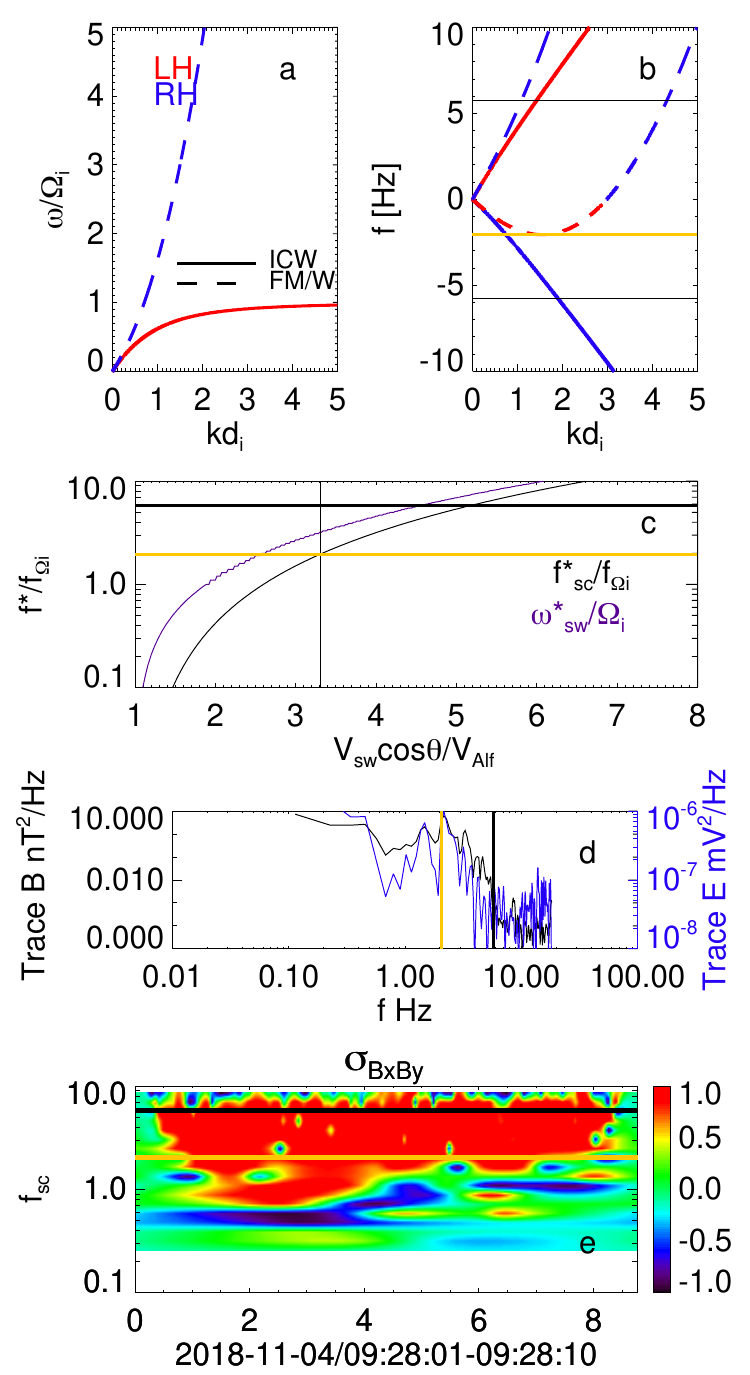}
    \caption{(a) ICW (solid) and FMW (dashed) cold-plasma dispersion curves in the solar wind frame. Red and Blue correspond to the left/right handed transverse wave polarization. (b) Doppler shifted cold plasma ICW and FMW dispersion into the spacecraft frame using measured $d_i$ and \Vsw; four curves are shown, outward propagating FMW and ICW (i,ii), and inward propagating FMW and ICW (iii,iv), each curve has been Doppler shifted outward by the solar wind. The minimum spacecraft frequency of the inward propagating FMW is shown in orange. For $|f|<f_{sc}^*$ there are at most six possible intersections of the Doppler shifted dispersion relations with $f_{sc}$. For $|f|>f^*_{sc}$ there are four possible intersections between $|f_{sc}|$ and the Doppler shifted Dispersion relation; of the four, only one is left hand polarized. (c) Shows the curve of $f_{sc}^*/f_{\Omega i}$ as a function of the ratio \vsw$\text{cos}\thetabv$/\va{ }for a parallel propagating wave; the corresponding curve in the plasma frame is shown in purple $\omega_{sc}^*/\Omega_i$. (d) Shows a coherent circularly polarized wave packet with left-hand polarization $|f|>f_{sc}^*$. (d) Wavelet spectrogram of circular polarization shows left-hand polarization  with $|f|>f_{sc}^*$ indicating that the wave must be an ICW.}
    \label{fig:f3}
\end{figure}

The cold plasma approximation, commonly used to model transverse electromagnetic plasma waves, has a dispersion-relation with two branches corresponding to different plasma-frame polarizations of the electric field relative to the mean background field:

\begin{equation}\left(\frac{\omega^\pm}{\Omega_i}\right)^2 =\left[ \frac{kd_i}{2}\left(
\sqrt{k^2d_i^2+4}\pm kd_i\right)\right]^2 \label{eq:disp}\end{equation}
where $\Omega_i=eB_0/m_i$ is the ion gyrofrequency and $d_i=\va/\Omega_i$ is the ion inertial length. The branch with $\ofm=\omega^+$ corresponds to the FMW wave with an intrinsic right-handed polarization, and the $\oicw=\omega^-$ branch corresponds to the left-handed ion-cyclotron wave; Figure \ref{fig:f3}(a) shows the two branches of the dispersion relation. Note that we take an approximation with $\omega <\Omega_{e}$, which is quite appropriate for the low-frequency waves under consideration.

In the spacecraft frame, these polarized transverse electromagnetic waves appear at the Doppler-shifted frequency

\begin{equation}
\label{eq:dop}
2\pi f= \omega(k) + \kv\cdot\Vsw.\end{equation}

For quasi-parallel propagation, characteristic of the transverse ion-scale waves measured in the solar wind \citep{Jian2010,Boardsen2015,Bowen2020a}, Equation \ref{eq:dop} reduces to \begin{equation} 2\pi f = \omega(k) \pm k\vsw cos\thetabv,
\end{equation}
where $\thetabv$ is the angle between the mean magnetic field and solar wind flow.

Observations from the solar wind in the inner-heliosphere and at 1 AU reveal a preference for occurrence of waves when the mean field is aligned with the solar wind flow direction, i.e. $\thetabv \sim 0$ or  $\thetabv \sim \pi$ \citep{Jian2010,Boardsen2015,Bale2019}. Few waves are observed when $\pi/4 <\thetabv< 3\pi/4$, though this is likely a sampling effect due to single spacecraft measurements of quasi-parallel waves in anisotropic turbulence \cite{Bowen2020a}.

The spacecraft frame frequency $f_{sc}$ (in units of Hz) is a positive definite quantity such that the observations at Dopper-shifted frequencies $f$, or $-f$, are both observed at $f_{sc}=|f|$. The plasma frame frequency, $\omega(k)$, is  intrinsically positive; however, the Doppler-shifted frequency of a wave to a negative value of $f$, causes a change of sign in the measured helicity in the Doppler-shifted (spacecraft) frame. 
For the two parallel propagating ICW and FMW modes, observation of a wave at the spacecraft frequency $f_{sc}$ most generally corresponds to one of eight cases 
\begin{align}
    \pm2\pi f_{sc}& =\oicw \pm k\vsw cos\thetabv \\
      \pm2\pi f_{sc}& =\ofm\pm k\vsw  \cos\thetabv,
\end{align}
where each $\pm$ sign on either side of the equations correspond to a set of two equations.

Several cases with no real solutions and can be discarded a priori: with the convention of positive frequency $\omega$ ($\bvec{k}\cdot\Vsw<0$ corresponds to backwards propagation), the addition of $k$ cannot produce a negative spacecraft frequency. Thus, outward propagating FMW and  ICW ($\bvec{k}\cdot\Vsw>0$) are never Doppler-shifted to negative frequency, and always appear in the spacecraft frame with their intrinsic plasma-frame polarization. Additionally, consideration of the ICW dispersion in the low $k$ limit gives a phase speed $\vph^{ICW}=\va$, with {\vph}  monotonically decreasing with increasing $k$. Under the condition $\va <\vsw$, inward propagating ICWs are observed at negative spacecraft frequencies--with an observed right-hand polarization in the spacecraft frame (we note that in future perihelion encounters, the $\va <\vsw$ ordering may not hold and a more general approach must be taken).

 Figure \ref{fig:f3}(b) shows the Doppler shifted cold plasma ICW and FMW dispersion equations with measured $\Omega_i$, $d_i,$ and \Vsw, from an interval with significant signature of left-hand polarization at 09:28 on November, 04 2018. For a wave at a given spacecraft frequency $f_{sc}$ there are five possible combinations of Doppler shifted wave-modes: outward (inward) propagating ICW appearing at a positive (negative) $f$. Outward propagating FMW waves appear at a positive $f$ and inward propagating FMW waves may appear at either positive or negative $f$. From Equation \ref{eq:dop}, the inward propagating FMW appears right-handed as long as its plasma frame frequency is larger than the Doppler shift, i.e.  $\ofm>|\kv\cdot\Vsw|$. The inward propagating FMW will occur as a left hand wave in the spacecraft frame at a frequency if $\ofm<|\kv\cdot{\Vsw}|$.

Clearly, measurements of the spacecraft-frame helicity of the magnetic field do not uniquely determine wave-mode and propagation direction. However, in some cases the cold-plasma dispersion can constrain wave-modes within a range of frequencies. Specifically, only two possible cases lead to left handed spacecraft-frame polarization, e.g. Figure\ref{fig:f3}(b): outward propagating ion cyclotron waves and inward propagating FMW waves with $|\kv\cdot{\Vsw}|>\ofm$. However, outside of the limit $k_{di} <<1$, \ofm grows faster than $k$ (e.g. Equation \ref{eq:disp}) while the Doppler shift term is proportionate with $-k$. Continuity of the dispersion relation imposes a critical wave number at which the FMW phase speed is greater than the solar wind speed, such that Doppler shift cannot change the sign of the handedness in the spacecraft frame. Thus a minimum value exists $f^{*}$, i.e. a maximum spacecraft frequency $f^*_{sc}=|f^{*}|$ at which, an inward propagating FMW can appear left-handed, e.g. Figure \ref{fig:f3}(b). Figure \ref{fig:f3}(c) shows the curve of $f^*_{sc}$ as a function of $\vsw\text{cos}\theta/\va$, determined through Equations \ref{eq:disp} and \ref{eq:dop}. A left-handed wave observed at $|f| > f^*_{sc}$ is out of the possible frequency range for an inward propagating, Doppler shifted, FMW and is thus uniquely explained by an outward propagating ICW.

\begin{figure}
    \centering
    \includegraphics{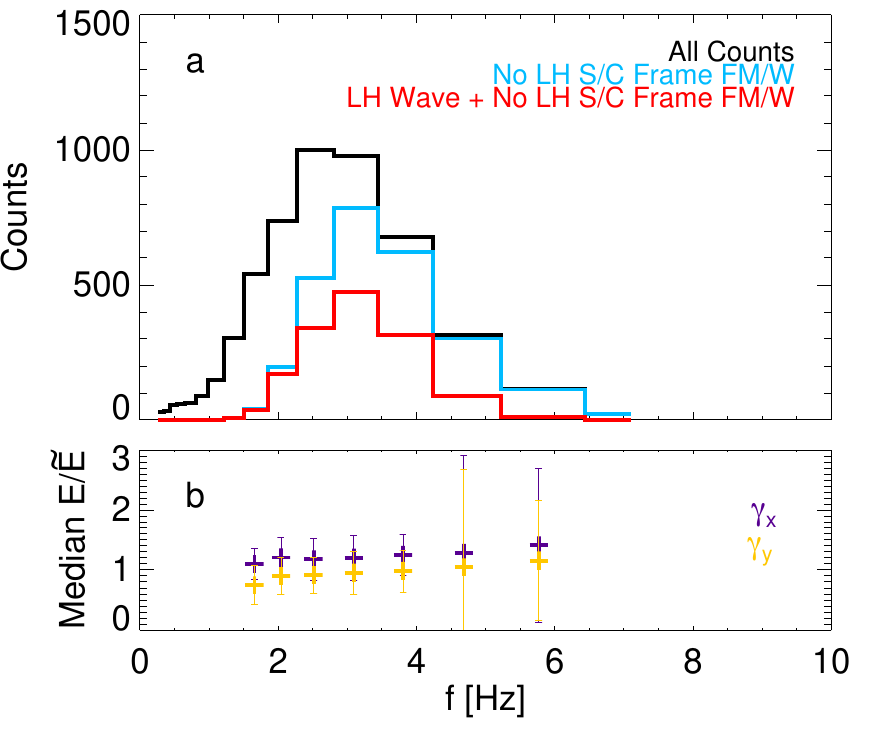}
    \caption{(a) Distribution of measured ion-scale waves on 04 November, 2018 (black). The subset of measured events which occur at a frequency $|f|>f_{sc}^{*}$ is shown in blue. Additionally, the  subset of waves with frequency $|f|>f_{sc}^{*}$ and have a left-handed polarization is shown in red. (b) Mean ratio between measured rms electric fields for left-handed waves with $|f|>f_{sc}^{*}$ and theoretical rms electric fields, assuming outward propagating ICW. At each wavelet scale; error bars show one standard deviation. The $x$ and $y$ antennas are shown in purple/orange.}
    \label{fig:f4}
\end{figure}

As a specific example, the curves in Figure \ref{fig:f3}(b) show Doppler shifted cold dispersion equations with measured $\Omega_i$, $d_i,$ and \Vsw. The minimum negative frequency for a FMW with left-handed spacecraft frame polarization is approximately 2 Hz. However, left-hand circularly polarized power extends up to $\sim6$ Hz Figure \ref{fig:f3}(d-e); at these frequencies the outward propagating ICW is the only permitted left-handed mode in the spacecraft frame. 

Left-handed waves with $|f|>f^{*}_{sc}$  allow for a direct comparison of FIELDS observations of \Esc, \B, and the derived {\Veff} with the values predicted by cold plasma dispersion. Figure \ref{fig:f4}(a) shows the number of circularly polarized measurements on Nov, 04 2019 at each wavelet scale. For each interval, the maximum spacecraft frequency that a FMW wave can appear left handed ($f^*_{sc}$) is determined using measured values of $d_i$, \Vsw, and the cold plasma dispersion equation (Equation \ref{eq:disp}). The distribution of all waves with $|f|>f^{*}_{sc}$ is shown in blue, and the distribution of left-hand polarized waves with $|f|>  f^*_{sc}$ is additionally shown in red--we take this as a measured distribution of outward propagating ICWs.

For each of these ICW events, the minimum variance direction of $\B$ is determined and taken as the wave propagation direction \citep{SonnerupCahill,Means1972,Santolik2003,Verniero2020}. The observation of $|f|>f^{*}_{sc}$ and left handed polarization breaks the degeneracy of the MVA determination, and the minimum eigenvector corresponding to outward propagation is chosen as $\kv$.  The cold plasma phase speed corresponding to the intersection of a Doppler shifted ICW with the wave frequency $f^{*}_{sc}$ is determined through a Newton-Raphson  root finding algorithm. Estimation of the phase speed allows construction of a synthetic solar wind frame electric field {\TEsw} obtained from {\B} as $${\TEsw}^{ICW}=-v_{ICW} {\hkv}\times\B.$$ Subsequently, a model spacecraft frame electric field is given as  

\begin{align}
{\TEsc}&={\TEsw}^{ICW} -\Vsw\times\B\\
{\TEsc}&=-(v_{ICW} {\hkv} +\Vsw)\times\B \\
{\TEsc}&=-{\TVeff}^{ICW}\times\B \label{eq:synthE}
\end{align}

Knowledge that waves with left hand polarization in the spacecraft frame with $|f|>f^{*}_{sc}$ must be on the ICW branch, allows for comparison of the theoretical {\TEsc} to the measured {\Esc}. Equation \ref{eq:synthE}, predicts the rms quantities measured in the spacecraft frame,

$$\tilde{V}_{z+}=\sqrt{\frac{\langle \tilde{E}_{scx}^2\rangle}{\langle{B_{scy}^2\rangle}}}$$
$$\tilde{V}_{z-}=\sqrt{\frac{\langle \tilde{E}_{scy}^2\rangle}{\langle{B_{scx}^2\rangle}}}.$$

Additionally, we test the frequency dependent effective length factors applied in \cite{Mozer2020} by comparing the theoretical ICW electric fields with the empirical measurements through calculation of

using \begin{align}
    \gamma_x=\sqrt{\frac{\langle {E}_x^2\rangle}{\langle \tilde{E}_x^2\rangle}}\\
    \gamma_y=\sqrt{\frac{\langle {E}_y^2\rangle}{\langle \tilde{E}_y^2\rangle}}\\
\end{align}

Figure \ref{fig:f4}(b) shows the mean $\gamma_x$ and $\gamma_y$ measured at each wavelet scale, with error bars given by one standard deviation; only scales with more than 10 counts were evaluated. The correction factor $\gamma_x$ is systematically larger than $\gamma_y$ indicating that measured electric fields are systematically larger in the $x$ direction than the $y$ direction. The larger uncertainties at higher frequencies are likely the result of measuring low amplitude polarized signals at the edge of broadband wave peak: \cite{Bowen2020a} show that the proton-gyroscale ($\sim$4-5 Hz) typically limits the distribution of circularly polarized waves. 

Incorporating $\gamma$ into the effective antenna lengths give corrections $$\alpha'_x=\alpha\gamma_x$$ and $$\alpha'_y=\alpha\gamma_y.$$ A scale-by-scale gain correction is then applied to the measured electric fields. As these correction factors are of order unity, their omission does not significantly change the subsequent results--or more importantly, our conclusions. 

\begin{figure*}
    \centering
    \includegraphics{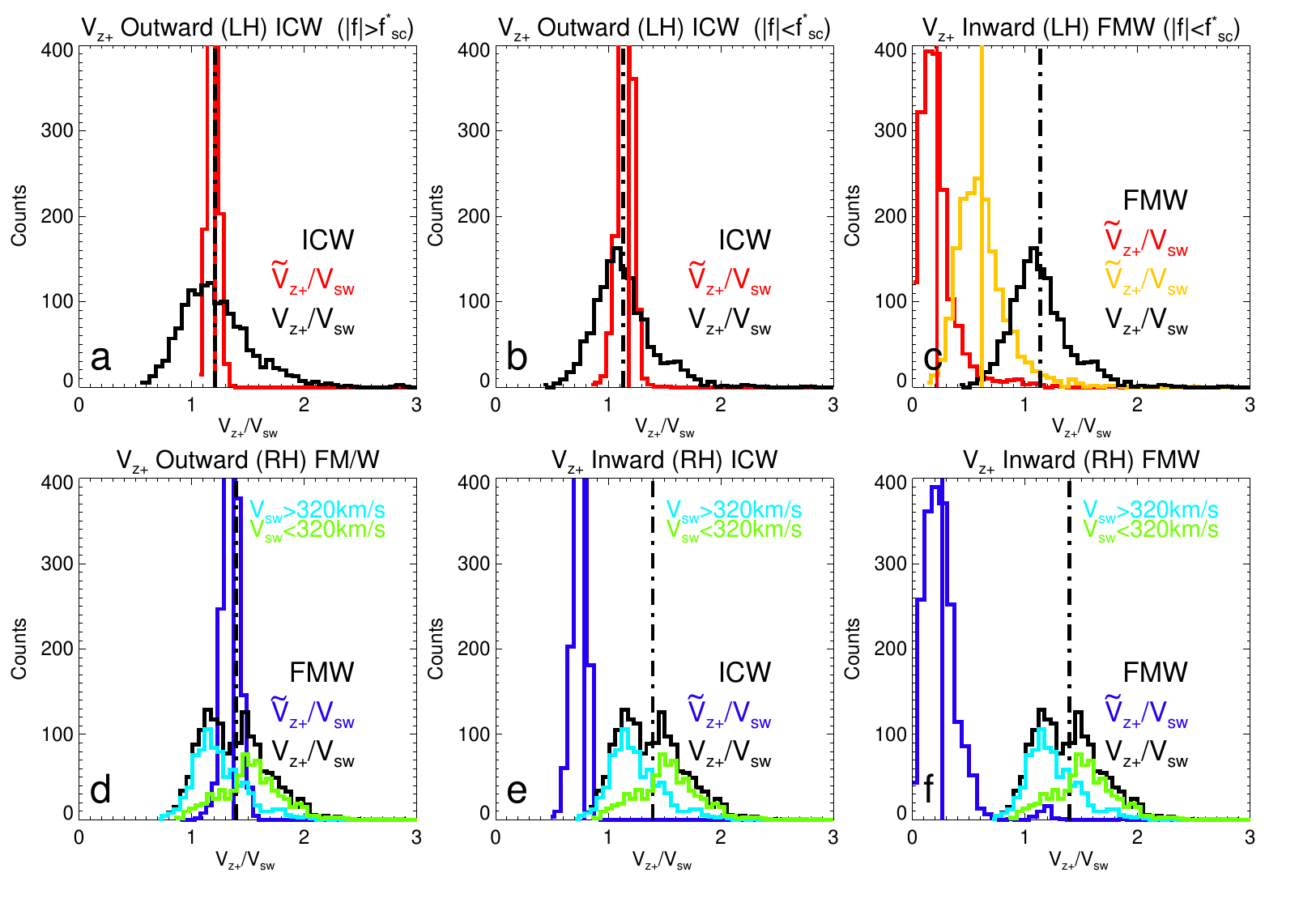}
    \caption{(a) Distributions of measured ${V}_{z+}$ (black) and synthetic $\tilde{V}_{z+}$ (red) normalized to $V_{sw}$ for left hand wave events with $|f|>f_{sc}^*$, where synthetic data is constructed for outward propagating ICW. (b) Measured (black) and synthetic (red) normalized to $V_{sw}$ for events with $|f|<f_{sc}^*$, where synthetic data is constructed for outward propagating ICW. (c) Measured and synthetic distributions for events with $|f|<f_{sc}^*$ where synthetic data is constructed for inward propagating FMW waves; two roots are possible for the synthetic cata (red and orange). (d) Distributions of measured ${V}_{z+}$ (black) and synthetic $\tilde{V}_{z+}$ (blue) normalized to $V_{sw}$ for right-hand events; synthetic data is constructed for outward propagating FM. (e) Distributions of measured (black) and synthetic (blue) data for right-hand events; synthetic data is constructed for inward propagating ICW. (f) Distributions of measured and synthetic data normalized for right-hand events with synthetic data corresponding to an inward propagating FMW. The measured distribution shown in (d-f) has two peaks, which correspond to two individual wind streams; the distribution is broken into data with $\vsw<320$ km/s (green) and  $\vsw>320$ km/s (teal)}
    \label{fig:f5}
\end{figure*}

 Application of $\gamma_x$ and $\gamma_y$ to waves with a left-hand spacecraft frame polarization enables measurements of rms speeds

$${V}_{z+}=\sqrt{\frac{\langle \tilde{E}_{scx}^2\rangle}{\langle{B_{scy}^2\rangle}}}$$
$${V}_{z-}=\sqrt{\frac{\langle \tilde{E}_{scy}^2\rangle}{\langle{B_{scx}^2\rangle}}}.$$
which constrains the mode and wavevector composition of the wave population. Figure \ref{fig:f5} shows distributions of synthetic $\tilde{V}_{z\pm}$ and measured $\tilde{V}_{z\pm}$, normalized to $\vsw$, with correction factors $\gamma_x$ and $\gamma_y$ applied; as the correction factors are of order unity, their omission does not significantly change the subsequent results. 

 Figure \ref{fig:f5}(a) shows the distribution of measured ${V}_{z+}/V_{sw}$ for waves with $|f|>f_{sc}^{*}$, along with the synthetic distribution $\tilde{V}_{z+}/V_{sw}$ constructed from the theoretical properties for outward propagating ICWs.  Figure \ref{fig:f5}(b-c) shows the distribution for left handed waves, but with  $|f|<f^{*}$, for which inward propagating FMW may appear left-handed. In Figure \ref{fig:f5}(b) we compare the measured distribution to synthetic quantities $\tilde{V}_{z+}/V_{sw}$ corresponding to outward propagating ICW; Figure \ref{fig:f5}(c) shows synthetic quantities corresponding to inward propagating FMW waves (two roots are generally available). Notably, the distribution of measured ${V}_{z+}/V_{sw}$ for frequencies $|f|<f^{*}$ is similar to the synthetic distribution of outward-propagating ICW.

Similar distributions for measured ${V}_{z-}/V_{sw}$ and synthetic $\tilde{V}_{z-}/V_{sw}$ are computed with identical results as presented for ${V}_{z+}/V_{sw}$ in Figure \ref{fig:f5}(a-c). Due to the redundancy in results producing an additional figure, we elect not to display both sets of distributions. Analysis of  ${V}_{z+}$ and ${V}_{z-}$ both strongly suggest that observed distribution of left-hand population of waves have an effective speed similar to outward propagating ICWs.

Figure \ref{fig:f5}(d-f) show measured distributions of ${V}_{z+}/V_{sw}$ for all observed right-handed waves. Synthetic distributions $\tilde{V}_{z+}/V_{sw}$ for outward propagating FMW waves are shown in Figure \ref{fig:f5}(d), while synthetic distributions for inward propagating ICW are shown in Figure \ref{fig:f5}(e). Synthetic data corresponding to an inward- propagating FM, which maintains right-hand polarization is shown in Figure \ref{fig:f5}(f). The measured distribution of data again largely has ${V}_{z+}/V_{sw}>1$, consistent with a dominant population of outward propagating FMW. Careful inspection of the measured distributions in Figure \ref{fig:f5}(d-f) reveals the presence of two peaks, which we identify as corresponding to two separate wind streams in the data. Separate populations are plotted for $\vsw>320$ km/s and $\vsw<320$ km/s. The synthetic data does not resolve two separate distributions. In either case, both  distributions are roughly consistent with outward propagating FMW waves. Interestingly, no bi-modality associated with stream-speed is observed in the left-handed waves. 

Again, data for ${V}_{z-}/V_{sw}$ were similarly analyzed. However, identical results are obtained as that from Figure \ref{fig:f5}(d-f), suggesting that the right-hand population of waves are propagating in the spacecraft frame with effective speeds similar to outward propagating FMW. To avoid redundancy, the distributions of ${V}_{z-}/V_{sw}$ are not shown.

\section{Discussion}
PSP observations of circularly polarized waves in the spacecraft frame suggest that effective wave speeds are typically larger than the bulk solar wind flow, indicating that most waves are likely outward propagating. While sub-dominant populations are present with effective speeds less than the solar wind speed, Figure \ref{fig:f5}(a-c), it is not clear whether this occurs as the result of statistical noise, or whether a subdominant population of inward propagating FMW is actually present. Further case studies may reveal the existence of individual inward propagating transverse waves. 

Outward propagating waves, which are Doppler-shifted to larger spacecraft frequencies by the solar wind, (with our convention of positive plasma frame frequency $\omega$), cannot occur at a negative frequency in the spacecraft frame. This suggests that, typically, no inversion of helicity occurs due to Dopper shift: e.g. left-hand ICWs observed in the spacecraft frame retain a left handed polarization. Thus, for the vast majority of measured waves, it is likely that the measured helicity corresponds to their intrinsic plasma-frame polarization. 

Doppler shifting the cold-plasma dispersion relation reveals a maximum frequency, $f^*_{sc}$ for which inward-propagating right-hand FMWs appear left handed in the spacecraft frame. Left-handed waves above this frequency are uniquely described by ICWs propagating outward. The measured electric field of waves which are left-handed and have $f>f^*_{sc}$, are to within our certainty, consistent with theoretical predictions for ICW made by the cold plasma dispersion. This analysis is made possible by an empirical verification of electric field calibration presented in \cite{Mozer2020} in the range of ion-scales. 

Many previous studies utilize the low wave-number approximation $kd_i\ll 1$, in which  $\vph\sim \va$, 
in order to study ion scale waves \citep{Jian2009,Jian2014,Gary2016,Wicks2016,Woodham2019,Bowen2020a}. However, the low wave number approximation is not broadly applicable, as these waves occur at ion-scales with $kd_i \sim 1$; our consideration of the full plasma dispersion presents a significant advancement in understanding ion-scale waves. \cite{Verniero2020} similarly pursued a more explicit estimate of $\mathbf{k}$, through Doppler-shifting the warm-plasma dispersion relation to determine the intersection with spacecraft frame frequency. The method of Doppler-shifting the dispersion relation curves is a promising method to infer intrinsic wave handedness, and has the advantage of being independent of the validity of the Taylor hypothesis.

The FMW and ICW dispersion relations are both non-linear at $kd_i \sim 1$, e.g.~ Figure \ref{fig:f3}, implying that the observed ion-scale waves should be subject to dispersive effects. However, broadband wave packets are commonly observed, which contain a range of wave-phase speeds, e.g. Figure \ref{fig:f3}(e). Over relatively short time-scales, broadband wave-packets at $\kdi \sim 1$ can disperse significantly.  The lack of observations of waves with dispersive characteristics suggests that waves are driven locally, and may be quickly damped. In contrast, if the waves are subject to WKB like transport, the dispersion of packets should be evident. It may be possible to constrain wave-packet life times, and thus damping, and associated heating rates, by thoroughly considering the dispersive effects on broadband waves.

We note that ICW and FMW wave phase speeds computed with warm-plasma dispersion relations will deviate from values derived under the cold plasma approximation. However, over the range of observed waves around $kd_i=[0.3,1]$, \citet{Verniero2020} show that for core $\beta=0.43$, beam-to-core drift speed of 1.5$\va$, and $\alpha$-to-core drift speed of 0.7$\va$, the ICW warm plasma phase speed is ~9\%-33\% lower than the ICW cold plasma phase speed; likewise, the FMW phase speed is ~1.3\%-3.6\% faster the FMW cold plasma phase speed (see Figure 7(c) of \citet{Verniero2020}). These perturbations are not significant enough to affect our interpretation of Figure \ref{fig:f5}; however, warm plasma effects may contribute to the dispersion in the measured distribution of ${V}_{z+}/V_{sw.}$ We suggest that while solutions to the warm-plasma dispersion are needed to understand growth and damping rates of these waves, propagation remains well described by the cold-plasma approximation.

Our measurements of effective phase speeds suggest that the vast majority of waves are outward propagating. Observations of left and right handed waves have at times been interpreted as observations of counter propagating waves, e.g. \cite{Jian2009}. The observed radial scaling of wave properties (e.g. amplitudes and occurrence rates) has been attributed to differences in transit time \citep{Jian2010,Boardsen2015}. We argue that radial scaling is more likely related to the heliospheric evolution of the ion distribution function \citep{Hellinger2013,Huang2020}.


Observations and theoretical analysis of quasi-parallel propagating ion-scale waves, suggest three general types of ion-driven instabilities that result in either left or right-hand wave generation \citep{Jian2009,Jian2010,Jian2014,Wicks2016,Gary2016,Verniero2020}: (1) proton anisotropies with $T_\perp > T_\parallel$ drive the ion-cyclotron instability \cite{Gary2001}, resulting in left-handed ICWs, (2) proton anisotropies with $T_\parallel > T_\perp$ drive the firehose instability, generating right-handed FMWs \cite{Hellinger2006}, and (3) relative drift speeds between ion components drive magnetosonic instabilities, also producing right-handed FMWs \citep{Goldstein2000,Gary2000,Gary2016}. Though \cite{PodestaGary2011b} suggest that FMW driven through temperature anisotropy should be inward propagating, observations of proton distribution functions at 1 au suggest that both temperature anisotropy and core-beam drift instabilities, are capable of driving outward propagating FMW \citep{Gary2016}.Furthermore, \cite{Verniero2020} show that waves observed by PSP are likely correlated with a proton beam. Our observation of a mixed distribution of outward propagating wave-modes suggests that a variety of instabilities are important in the young solar wind.

Generally, \cite{Klein2018} suggest that the solar wind at 1 au is unstable approximately half of the time, with a vast range of both fluid and kinetic scale instabilities capable of converting free energy into electromagnetic waves \citep{Yoon2017,Verscharen2019}. In the inner-heliosphere, these instabilities are ever-more present \citep{Klein2019}. Our constraint of mode composition and wave vector distribution of ion scale waves in the solar wind through electric field measurements, provides a key step in understanding how energy in the wave-field is redistributed to the plasma.

\section{Acknowledgements}
The FIELDS instrument was designed and developed under NASA contract NNN06AA01C.

\end{document}